 \documentclass[final,3p,twocolumn,10pt]{elsarticle}

\usepackage{epsfig}

\usepackage{amssymb}
\usepackage{amsmath}

\usepackage{dcolumn}

\biboptions{numbers,super,sort&compress,longnamesfirst}

\journal{Physica A}

\begin{document}

\begin{frontmatter}



\title{The influence of the liquid slab thickness on the planar vapor-liquid interfacial tension}


\author{Stephan Werth, Sergey Lishchuk, Martin Horsch,$^1$ Hans Hasse}


\address{Laboratory of Engineering Thermodynamics, Department of Mechanical and Process Engineering, University of Kaiserslautern, Erwin-Schr\"odinger-Str. 44, 67663 Kaiserslautern, Germany}

\begin{abstract}
One of the long standing challenges in molecular simulation is the description of interfaces. On the molecular length scale, finite size effects significantly influence the properties of the interface such as its interfacial tension, which can be reliably investigated by molecular dynamics simulation of planar vapor-liquid interfaces. For the Lennard-Jones fluid, finite size effects are examined here by varying the thickness of the liquid slab. It is found that the surface tension and density in the center of the liquid region decreases significantly for thin slabs. The influence of the slab thickness on both the liquid density and the surface tension is found to scale with 1/$S^3$ in terms of the slab thickness $S$, and a linear correlation between both effects is obtained. The results corroborate the analysis of Malijevsk\'y and Jackson, J.\ Phys.: Cond.\ Mat.\ 24: 464121 (2012), who recently detected an analogous effect for the surface tension of liquid nanodroplets.
\end{abstract}

\begin{keyword}
Surface tension \sep long range correction \sep finite size effects

\end{keyword}

\end{frontmatter}


\section{Introduction}
\label{}
\footnotetext[1]{Corresponding author: Martin Thomas Horsch, martin.horsch@mv.uni-kl.de, +49 631 205 4028.}


\noindent
Molecular simulation is a well-established approach for the analysis of fluid interfaces and their molecular structure. 
Much work has been dedicated to the interfacial tension.\cite{YYF10,TGWCR84,BB92,OVBBKS12}\
For a fluid interface, there are (at least) three different aspects in which its size can be varied, each of which may affect the interfacial tension:
\begin{itemize}
   \item{} curvature effects, depending on the local characteristic radii of curvature
   \item{} capillary wave effects, depending on the range of wavelengths permitted by the morphology and size of the interface
   \item{} confinement effects, which arise due to spatial restrictions imposed on a fluid phase by one or several interfaces or walls
\end{itemize}
According to the Tolman \cite{Tolman49} approach, the interfacial tension of a nanodroplet deviates from that of a planar interface due to its extremely curved shape.\cite{HB94,HHSAEVMUJ12,ZY11,TOBVB12,Buff51}\ However, it should be noted that all three phenomena are present when the size of a droplet is varied: Smaller droplets have a higher curvature, a smaller range of capillary wavelengths, and a more significant deviation from bulk-like behaviour due to confinement. In addition to curvature, the other effects might therefore also have a significant influence on the formation of droplets in a supersaturated vapor, where the size of the critical nucleus and the nucleation rate are of major interest.\cite{Lovett07,VHH09,Kraska06,CWSWR09,BKA08}\ A similar case is cavitation, where bubbles emerge in a liquid phase.\cite{BC00,SBB06,EMB03}

\begin{figure}[b!]
\centering
 \includegraphics[width=7.777cm]{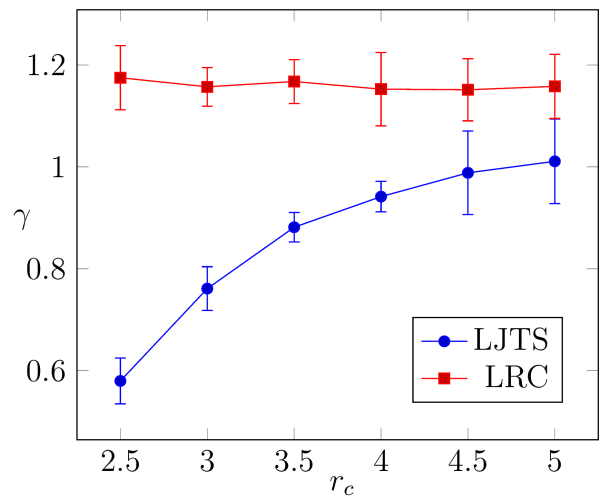}
\caption{Dependence of the computed surface tension $\gamma$ on the employed cutoff radius $r_\mathrm{c}$, for the truncated and shifted LJ potential without long-range effects (LJTS) and for the full LJ potential with a long-range correction (LRC) according to Jane\v{c}ek.\cite{Janecek06}\ The simulations were carried out in the canonical ensemble with $N$ = 2 048 particles at $T$ = 0.7 using a simulation box with an elongation of $\ell_y$ = 66 in the direction perpendicular to the vapor-liquid interfaces.}
\label{fig:rc}
\end{figure}

Spherically curved interfaces of droplets were simulated for the first time by molecular simulation in the early 1970s.\cite{McGinty73,LBA73}\ Nonetheless, while curvature effects are relatively well-studied, there are only astonishingly few investigations related to the other size effects, which are also present in the case of planar interfaces. Several previous works address the influence of small simulation volumes,\cite{OLA05, GJBM05, Janecek09, WSMB97}\ which is usually discussed in terms of capillary wave effects.\cite{FRLP66, LVFSHBIA08, SL89, GF90, WSMB97}\ The present study considers the influence of the liquid slab thickness, i.e.\ of confinement by two parallel planar vapor-liquid interfaces which are close to each other. This effect was previously investigated by Weng et al.\ \cite{WPLT00} who, however, did not find a systematic correlation.

The computation of interfacial properties is always done in a single simulation volume containing both phases, the liquid and the vapor phase, separated by the interface. For the calculation of the bulk properties there are many other methods, like Grand Equilibrium,\cite{VH02}\ $NpT$ plus test particle,\cite{MF90}\ or the Gibbs ensemble method.\cite{Pana02}\ The surface tension can be computed for example via the virial route or the surface free energy.\cite{SM91,TGWCR84,GM12,WIH12,HKT09,Binder03}\ The virial route is directly related to the common approach for calculating the pressure in a molecular simulation.\cite{LA12,MBL97,GJBM05,Janecek09}\ It is known that the pressure in dependence of the density exhibits a van der Waals loop in the two phase region.\cite{IK09,KRI09}

\begin{figure}[h!]
\centering
 \includegraphics[width=7.777cm]{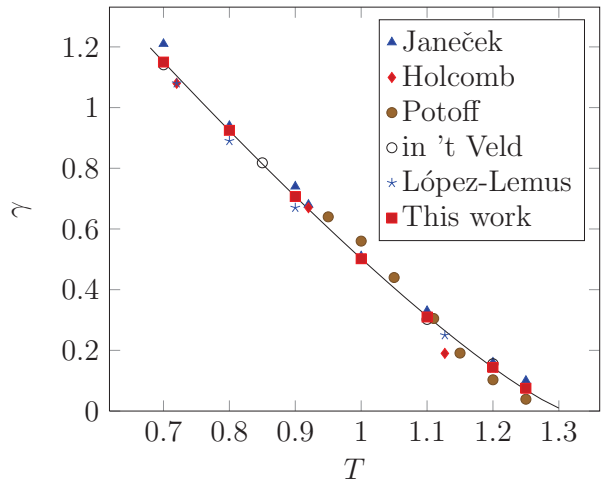}
\caption{Surface tension $\gamma$ over the temperature for large liquid slabs. Comparison of results of different authors: (red squares) This work, (blue triangles) Jane{\v{c}}ek,\cite{Janecek06}\ (red diamonds) Holcomb et al.,\cite{HCZ93}\ (brown circles) Potoff and Panagiotopoulos,\cite{PP00}\ (black circles) in't Veld et al.,\cite{VIG07}\ (blue stars), as well as L\'opez Lemus and Alejandre,\cite{LA02}; (solid line) regression from Eq.\ (\ref{eq:surface}).}
\label{fig:s_t}
\end{figure}

At interfaces, the long range contribution to the interaction potential plays an important role for all thermodynamic properties.\cite{AC10,MT91,GBNKBM12}\ Nonetheless, there are also simulations applying a truncated and shifted potential, which neglects the whole long range contribution.\cite{BC00,GB09,VKFH06,GKBSR07}\ When dealing with a homogeneous system, long range corrections are only needed for energy and pressure,\cite{Neumann85,Onsager36}\ while in an inhomogeneous configuration, also the dynamics of the systems needs to be appropriately corrected.\cite{Janecek06,LVF90,MWF97}

\begin{figure}[t!]
\centering
 \includegraphics[width=7.777cm]{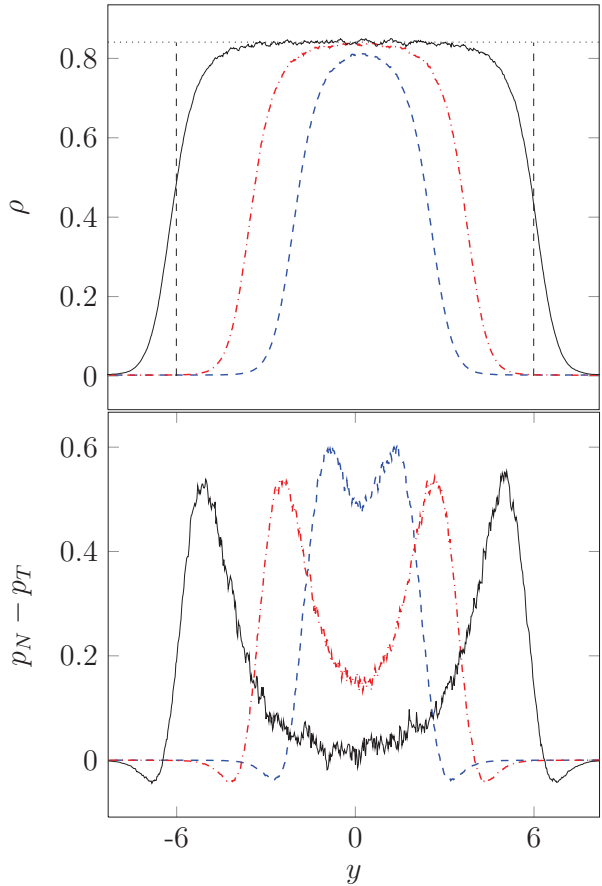}
\caption{Density $\rho$ (top) and differential pressure $p_N - p_T$ (bottom) over the $y$ coordinate (i.e.\ the direction perpendicular to the interface). The temperature is $T = 0.7$. The blue dashed line corresponds to the minimum stable configuration which is $S = 4.3$ for this temperature, while the red dash dotted one corresponds to $S$ = 7 and the black solid one to $S$ = 12. The dotted line in the upper picture represents the bulk liquid density and the difference between the vertical dashed lines in the upper picture represent the equimolar slab thickness.}
\label{fig:ddy}
\end{figure}

In the present work, the influence of the liquid slab thickness on thermodynamic properties is discussed. A suitable long range correction is used to obtain accurate and validated results. 

\section{Simulation method}

\begin{figure}[b!]
\centering
 \includegraphics[width=7.777cm]{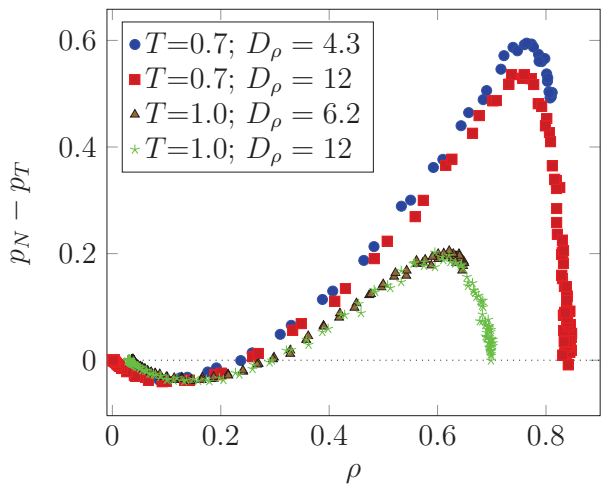}
\caption{Differential pressure $p_N - p_T$ over the density for the temperature $T = 0.7$ and $T = 1.0$. The blue circles and brown triangles correspond to the minimum stable configuration for the corresponding temperature, while the red squares and green stars correspond to $S$ = 12. The dotted line represents the zero line.}
\label{fig:vdW_07}
\end{figure}

\noindent
In this study, the Lennard-Jones potential (LJ) 
\begin{equation}
 u_{ij} = 4 \epsilon \left[ \left( \frac{\sigma}{r_{ij}} \right)^{12} - \left( \frac{\sigma}{r_{ij}} \right)^{6}\right]
\end{equation}
is employed, where $\epsilon$ and $\sigma$ are the energy and size parameters and $r_{ij}$ is the distance between the two particles $i$ and $j$. The standard LJ para\-meters $\epsilon=1$ and $\sigma=1$ are used, yielding a reduced LJ potential.

\begin{table}[h!]
\centering
\begin{tabular}{c|ccc||cc}
   $T$ & $N$ & $\ell_y$ & $\ell_\parallel$ & $S$ & $\gamma$ \\ \hline

   $0.7$ & $300$ $000$ & $100\phantom{.0}$ & $94.2$ & $40.0\phantom{0}$ & $1.150(4)$ \\
      & $\phantom{0}16$ $000$ & $\phantom{0}50.0$ & $39.7$ & $12.0\phantom{0}$ & $1.14(7)\phantom{0}$ \\
      & $\phantom{0}16$ $000$ & $\phantom{0}50.0$ & $45.8$ & $\phantom{0}9.0\phantom{0}$ & $1.14(6)\phantom{0}$ \\
      & $\phantom{0}16$ $000$ & $\phantom{0}50.0$ & $51.8$ & $\phantom{0}7.0\phantom{0}$ & $1.13(3)\phantom{0}$ \\
      & $\phantom{0}16$ $000$ & $\phantom{0}50.0$ & $61.1$ & $\phantom{0}5.0\phantom{0}$ & $1.10(2)\phantom{0}$ \\
      & $\phantom{0}16$ $000$ & $\phantom{0}50.0$ & $64.3$ & $\phantom{0}4.5\phantom{0}$ & $1.08(1)\phantom{0}$ \\
      & $\phantom{0}16$ $000$ & $\phantom{0}50.0$ & $65.7$ & $\phantom{0}4.3\phantom{0}$ & $1.06(2)\phantom{0}$ \\ \hline

   $0.8$ & $300$ $000$ & $100\phantom{.0}$ & $96.4$ & $40.0\phantom{0}$ & $0.93(1)\phantom{0}$ \\
      & $\phantom{0}16$ $000$ & $\phantom{0}50.0$ & $40.3$ & $12.0\phantom{0}$ & $0.92(6)\phantom{0}$ \\
      & $\phantom{0}16$ $000$ & $\phantom{0}50.0$ & $46.4$ & $\phantom{0}9.0\phantom{0}$ & $0.92(3)\phantom{0}$ \\
      & $\phantom{0}16$ $000$ & $\phantom{0}50.0$ & $52.3$ & $\phantom{0}7.0\phantom{0}$ & $0.91(3)\phantom{0}$ \\
      & $\phantom{0}16$ $000$ & $\phantom{0}50.0$ & $61.2$ & $\phantom{0}5.0\phantom{0}$ & $0.87(3)\phantom{0}$ \\
      & $\phantom{0}16$ $000$ & $\phantom{0}50.0$ & $63.9$ & $\phantom{0}4.55$ & $0.85(2)\phantom{0}$ \\ \hline

   $0.9$ & $300$ $000$ & $100\phantom{.0}$ & $98.2$ & $40.0\phantom{0}$ & $0.707(8)$ \\
      & $\phantom{0}16$ $000$ & $\phantom{0}50.0$ & $40.8$ & $12.0\phantom{0}$ & $0.71(5)\phantom{0}$ \\
      & $\phantom{0}16$ $000$ & $\phantom{0}50.0$ & $46.5$ & $\phantom{0}9.0\phantom{0}$ & $0.70(4)\phantom{0}$ \\
      & $\phantom{0}16$ $000$ & $\phantom{0}50.0$ & $51.9$ & $\phantom{0}7.0\phantom{0}$ & $0.69(3)\phantom{0}$ \\
      & $\phantom{0}16$ $000$ & $\phantom{0}50.0$ & $55.4$ & $\phantom{0}6.0\phantom{0}$ & $0.68(2)\phantom{0}$ \\
      & $\phantom{0}16$ $000$ & $\phantom{0}50.0$ & $58.1$ & $\phantom{0}5.4\phantom{0}$ & $0.66(3)\phantom{0}$ \\
\end{tabular}
\caption{Surface tension $\gamma$ in dependence of the slab thickness $S$ (low temperatures). The total elongation of the simulation box is indicated as $\ell_y$ in the direction perpendicular to the vapor-liquid interfaces and as $\ell_\parallel$ ($= \ell_x = \ell_z$) in the other spatial directions. The statistical error in terms of the final digit is shown in parentheses.}
\label{tab:data_gamma1}
\end{table}

As usual, the potential was truncated in order to reduce the computing time. A cutoff radius of $r_c = 3$ was used for the present simulations. To diminish the error made by this assumption, a bin based tail correction was applied to the simulation.\cite{Janecek06, SME07}\ Thereby, the potential energy, the forces acting on the molecules, and the virial are each split into an explicitly computed part and a long range correction. The calculation of the correction terms was conducted every 10 time steps. It is known that this method provides cutoff independent results for the LJ fluid.\cite{Janecek06}\ For a discussion of the employed method in full detail, the reader is referred to Jane\v{c}ek's work,\cite{Janecek06}\ wherein this approach was first presented.

The interfacial tension $\gamma$ is given by the difference between the diagonal components of the virial tensor $\Pi_N-\Pi_T$ or, equivalently, an integral over the differential pressure\cite{Janecek06,WTRH83} $p_N - p_T$
\begin{equation}
 \gamma = \frac{1}{2 A} \left( \Pi_N - \Pi_T \right) = \int_{-\infty}^{\infty} \left( p_N - p_T \right) \text{d}y,
\end{equation}
where $2A$ denotes the interfacial area of the two interfaces. The pressure calculation is based on the method proposed by Irving and Kirkwood,\cite{IK49}\ but in contrast to their approach the pressure is not homogeneously distributed between the positions of the two particles. To speed up the simulation the pressure is divided between the bins of the involved particles. It should be noted that this modification has a slight effect on the localized pressure tensor but leads to the same outcome for the overall surface tension.\cite{Janecek09}

\begin{table}[b!]
\centering
\begin{tabular}{c|ccc||cc}
   $T$ & $N$ & $\ell_y$ & $\ell_\parallel$ & $S$ & $\gamma$ \\ \hline

   $1.0\phantom{0}$ & $300$ $000$ & $100\phantom{.0}$ & $100\phantom{.0}$ & $40.0\phantom{0}$ & $0.502(5)$ \\
      & $\phantom{0}16$ $000$ & $\phantom{0}50.0$ & $\phantom{0}40.1$ & $12.0\phantom{0}$ & $0.50(6)\phantom{0}$ \\
      & $\phantom{0}16$ $000$ & $\phantom{0}50.0$ & $\phantom{0}46.1$ & $\phantom{0}9.0\phantom{0}$ & $0.50(2)\phantom{0}$ \\
      & $\phantom{0}16$ $000$ & $\phantom{0}50.0$ & $\phantom{0}50.8$ & $\phantom{0}7.0\phantom{0}$ & $0.48(5)\phantom{0}$ \\
      & $\phantom{0}16$ $000$ & $\phantom{0}50.0$ & $\phantom{0}53.0$ & $\phantom{0}6.2\phantom{0}$ & $0.5(1)\phantom{00}$ \\ \hline

   $1.1\phantom{0}$ & $300$ $000$ & $100\phantom{.0}$ & $102\phantom{.0}$ & $40.0\phantom{0}$ & $0.310(4)$ \\
      & $\phantom{0}16$ $000$ & $\phantom{0}50.0$ & $\phantom{0}40.9$ & $12.0\phantom{0}$ & $0.31(4)\phantom{0}$ \\
      & $\phantom{0}16$ $000$ & $\phantom{0}50.0$ & $\phantom{0}45.3$ & $\phantom{0}9.0\phantom{0}$ & $0.30(3)\phantom{0}$ \\
      & $\phantom{0}16$ $000$ & $\phantom{0}50.0$ & $\phantom{0}48.6$ & $\phantom{0}7.25$ & $0.28(4)\phantom{0}$ \\
      & $\phantom{0}16$ $000$ & $\phantom{0}50.0$ & $\phantom{0}50.2$ & $\phantom{0}6.5\phantom{0}$ & $0.26(3)\phantom{0}$ \\ \hline

   $1.2\phantom{0}$ & $300$ $000$ & $100\phantom{.0}$ & $101\phantom{.0}$ & $40.0\phantom{0}$ & $0.144(8)$ \\
      & $\phantom{0}16$ $000$ & $\phantom{0}50.0$ & $\phantom{0}38.2$ & $12.0\phantom{0}$ & $0.14(5)\phantom{0}$ \\
      & $\phantom{0}16$ $000$ & $\phantom{0}50.0$ & $\phantom{0}40.8$ & $\phantom{0}9.0\phantom{0}$ & $0.14(4)\phantom{0}$ \\
      & $\phantom{0}16$ $000$ & $\phantom{0}50.0$ & $\phantom{0}42.4$ & $\phantom{0}7.5\phantom{0}$ & $0.13(4)\phantom{0}$ \\ \hline

   $1.25$ & $300$ $000$ & $100\phantom{.0}$ & $101\phantom{.0}$ & $40.0\phantom{0}$ & $0.075(4)$ \\
      & $\phantom{0}16$ $000$ & $\phantom{0}50.0$ & $\phantom{0}37.2$ & $12.0\phantom{0}$ & $0.08(5)\phantom{0}$ \\
      & $\phantom{0}16$ $000$ & $\phantom{0}50.0$ & $\phantom{0}39.2$ & $\phantom{0}9.0\phantom{0}$ & $0.06(2)\phantom{0}$ \\
      & $\phantom{0}16$ $000$ & $\phantom{0}50.0$ & $\phantom{0}39.9$ & $\phantom{0}8.0\phantom{0}$ & $0.06(4)\phantom{0}$ \\
\end{tabular}
\caption{Surface tension $\gamma$ in dependence of the slab thickness $S$ (high temperatures), cf.\ Tab.\ \ref{tab:data_gamma1}.}
\label{tab:data_gamma2}
\end{table}

Molecular dynamics (MD) simulations were conducted in the canonical ensemble with $N$ = 16 000 particles. The equimolar thickness of the li\-quid slab $S$ was varied between 12 and the minimum stable configuration. The equimolar thickness was determined using the saturated densities $\rho'$ and $\rho''$ for the given temperature, the simulation volume $V$, and the number of particles $N$
\begin{equation}
 S = \frac{N-\rho''V}{(\rho' - \rho'')A},
\end{equation}
i.e.\ $S$ only depends on the boundary conditions applied to the molecular simulation within the canonical ensemble, not on the outcome of the simulation, and it does not vary over simulation time. The temperature was kept constant by a velocity scaling thermostat. All simulations were performed in a parallelepiped box. The elongation of the simulation volume in $y$ direction, i.e.\ normal to the interface, was $\ell_y = 50$.

\begin{table}[h!]
\centering
 \begin{tabular}{l|D{.}{.}{2.6} D{.}{.}{2.6} D{.}{.}{2.6} D{.}{.}{2.6} }
      & & \multicolumn{1}{r}{$T$} & & \\
     $S$  	& 0.7 	& 0.8	&	0.9	& 1.0 	 \\
\hline
  	&		&	 &	&		\\
  12	&-0.01 \, (6)		 &-0.01 \, (4)	 	 &0.00  \,(3)		&0.00  \,(2)	\\
 \; 9	&0.04 \, (6)		 &0.04 \, (4) 		 &0.04  \,(2) 		&0.04  \,(5) 	\\
 \; 7	&0.13 \, (5)		 &0.12 \, (3) 		 &0.12  \,(4) 		&0.11  \,(5)	\\
 \; 6.2	&\multicolumn{1}{c}{n/a} &\multicolumn{1}{c}{n/a}&\multicolumn{1}{c}{n/a}&0.16  \,(5) 	\\
 \; 6	&\multicolumn{1}{c}{n/a} &\multicolumn{1}{c}{n/a}&0.18  \,(2) 		&\multicolumn{1}{c}{$\star$}	\\
 \; 5.4	&\multicolumn{1}{c}{n/a} &\multicolumn{1}{c}{n/a}&0.24  \,(3) 		&\multicolumn{1}{c}{$\star$}	\\
 \; 5	&0.34 \, (2) 	 	 &0.32 \, (3) 		 &\multicolumn{1}{c}{$\star$}	&\multicolumn{1}{c}{$\star$}	\\
 \; 4.55&\multicolumn{1}{c}{n/a} &0.38 \, (6) 		 &\multicolumn{1}{c}{$\star$}	&\multicolumn{1}{c}{$\star$}	\\
 \; 4.5 &0.43 \, (4) 		 &\multicolumn{1}{c}{$\star$}  &\multicolumn{1}{c}{$\star$} &\multicolumn{1}{c}{$\star$}	\\
 \; 4.3	&0.48 \, (3) 		 &\multicolumn{1}{c}{$\star$}  &\multicolumn{1}{c}{$\star$} &\multicolumn{1}{c}{$\star$}	\\
 \end{tabular}
\caption{Differential pressure $p_N - p_T$ in the center of the liquid slab in dependence on the slab thickness $S$ (low temperatures). ``n/a'': data not determined; asterisks: liquid slab found to be unstable.}
\label{tab:data_pd1}
\end{table}

For the simulation of a reference case the number of particles $N$ was increased to 300 000, the elongation in $y$ direction was $\ell_y = 100$ and a slab thickness of $S = 40$ was used. The temperature $T$ was varied between 0.7 and 1.25, i.e.\ from the triple point temperature to 95$\%$ of the critical temperature. The simulations were carried out using the $ls1$ molecular dynamics code.\cite{BBV11}\ The equation of motion was solved by a leapfrog integrator.\cite{Fincham92}\ A time step of $\Delta t=0.002$ was used. The equilibration was conducted for at least 120 000 time steps, while the production ran 
for 840 000 time steps. The statistical errors given in the present study are equal to 3 times the standard deviation of 7 block averages, each over 120 000 time steps. 

A further series of simulations was conducted to validate whether relatively small cutoff radii are permissible when Jane\v{c}ek's cutoff correction is employed. The simulation results support this conclusion, cf.\ Fig.\ \ref{fig:rc}.

\begin{table}[t!]
\centering
 \begin{tabular}{l|D{.}{.}{2.6} D{.}{.}{2.6} D{.}{.}{2.6}}
      & & T &  \\
     $S$  	& 1.1 	& 1.2	&	1.25	 \\
\hline
  	&		&	&			\\
  12	&0.01 \, (2)		&0.00 \, (5) 		&0.00 \,(3) 	\\
 \; 9	&0.05 \, (2) 		&0.02 \, (4) 		&0.01 \, (3) 	\\
 \; 8   &\multicolumn{1}{c}{n/a}&\multicolumn{1}{c}{n/a}&0.02 \, (2) 	\\
 \; 7.5 &\multicolumn{1}{c}{n/a}&0.03 \, (3) 		&\multicolumn{1}{c}{$\star$}			\\
 \; 7.25&0.10 \, (3) 		&\multicolumn{1}{c}{$\star$}	&\multicolumn{1}{c}{$\star$}			\\
 \; 6.5	&0.12 \, (2)		&\multicolumn{1}{c}{$\star$}	&\multicolumn{1}{c}{$\star$}			\\
 \end{tabular}
\caption{Differential pressure $p_N - p_T$ in the center of the liquid slab (high temperatures), cf.\ Tab.\ \ref{tab:data_pd1}.}
\label{tab:data_pd2}
\end{table}

\section{Results}

\noindent
As described above, a series of simulations was carried out using a large liquid slab of $S = 40$ in order to approximate bulk phase behavior. The resulting values $\gamma_\infty$ and $\rho'_\infty$ are used as a reference for the further simulations. The resulting surface tension is shown in Fig.\ \ref{fig:s_t}. The regression
\begin{equation}
\gamma = 2.94 \left( 1 - \frac{T}{T_c} \right) ^ {1.23} 
\label{eq:surface}
\end{equation}
is obtained, with $T_c = 1.3126$ according to P\'erez Pellitero et al.\ \cite{PPUOM06} The type of correlation is the same as proposed by Vrabec et al.\ \cite{VKFH06} for the truncated and shifted LJ potential, and their exponent (i.e.\ 1.21) is also very similar to the present one.

\begin{table}[p!]
\centering
 \begin{tabular}{l|llll}
      & & \multicolumn{1}{r}{$T$}& & \\
     $S$  	& 0.7 	& 0.8	&	0.9	& 1.0 	 \\
\hline
  	&		&	&	&		\\
  40	&0.8410 (2) 	&0.7974 (3) 	&0.7507 (6) 	&0.699 (1) 	\\
  12	&0.84 (3) 	&0.80 (2) 	&0.75 (1) 	&0.70 (1) 	\\
 \; 9	&0.84 (1) 	&0.80 (1) 	&0.75 (1) 	&0.69 (2) 	\\
 \; 7	&0.83 (1) 	&0.79 (1) 	&0.74 (2) 	&0.68 (2) 	\\
 \; 6.2	&\quad n/a	&\quad n/a	&\quad n/a	&0.66 (5) 	\\
 \; 6	&\quad n/a	&\quad n/a	&0.73 (2)	&\quad $\star$	\\
 \; 5.4	&\quad n/a	&\quad n/a	&0.72 (1) 	&\quad $\star$	\\
 \; 5	&0.82 (1) 	&0.77 (1) 	&\quad $\star$	&\quad $\star$\\
 \; 4.55&\quad n/a	&0.76 (3) 	&\quad $\star$	&\quad $\star$\\
 \; 4.5 &0.81 (1)	&\quad $\star$	&\quad $\star$	&\quad $\star$\\
 \; 4.3	&0.81 (1) 	&\quad $\star$	&\quad $\star$	&\quad $\star$\\
 \end{tabular}
\caption{Density $\rho'$ in the center of the liquid slab in dependence on the slab thickness $S$ (low temperatures).}
\label{tab:data_rho1}
\end{table}

Moreover, simulations were also performed for smaller slab thicknesses ($S \leq 12$). Thereby, MD runs were conducted with successively smaller values of $S$, until a minimum stable thickness was reached for the given temperature.

In Fig.\ \ref{fig:ddy}, the density and the corresponding differential pressure profile for $T=0.7$ is plotted over the $y$ coordinate. It can be seen that the density in the center of the slab for $S = 12$ and $S = 7$ almost matches the bulk liquid density at saturation, which is also plotted in Fig.\ \ref{fig:ddy} as reference. It is slightly smaller for $S = 4.3$. In the differential pressure the difference between the three simulations is more significant. The differential pressure in the center of the slab ($y=0$) almost reaches the zero line for $S = 12$, while for $S = 7$ and $S = 4.3$ the pressure tensor is anisotropic throughout the liquid slab. The differential pressure can be seen as an indicator for the fluid to be isotropic or homogeneous, i.e.\ not influenced by the two interfaces. For $S$ larger than 12, the differential pressure fluctuates around zero in the center of the liquid slab for $T = 0.7$.

\begin{table}[b!]
\centering
 \begin{tabular}{l|llll}
      & & $T$ & & \\
     $S$  	& 1.1 	& 1.2	&	1.25	 \\
\hline
  	&		&	&			\\
  40	&0.6393 (4)	&0.564 (2)	&0.515 (7) 	\\
  12	&0.63 (3)	&0.56 (3) 	&0.49 (10)	\\
 \; 9	&0.62 (2)	&0.54 (3) 	&0.46 (4) 	\\
 \; 8   &\quad n/a	&\quad n/a	&0.43 (15)	\\
 \; 7.5 &\quad n/a	&0.51 (7) 	&\quad $\star$		\\
 \; 7.25&0.59 (3)	&\quad $\star$	&\quad $\star$		\\
 \; 6.5	&0.57 (8)	&\quad $\star$	&\quad $\star$		\\
 \end{tabular}
\caption{Density $\rho'$ in the center of the liquid slab (high temperatures), cf.\ Tab.\ \ref{tab:data_rho1}.}
\label{tab:data_rho2}
\end{table}

In Fig.\ \ref{fig:vdW_07}, the plot of the differential pressure over the density is shown, which is obtained from the data displayed in Fig.\ \ref{fig:ddy}. Additionally results for $T = 1.0$ are shown. The plot in Fig.\ \ref{fig:vdW_07} exhibits van der Waals loops in all cases. The red squares and green stars correspond to a large liquid slab, while the blue circles and brown triangles show the result of the smallest stable liquid slab. Like in Fig.\ \ref{fig:ddy}, it is obvious that the differential pressure does not reach zero in the latter case.

The resulting surface tensions are shown in Tabs.\ \ref{tab:data_gamma1} and \ref{tab:data_gamma2}. For all temperatures the surface tension decreases when the liquid slab thickness decreases.

\begin{figure}[h!]
\centering
 \includegraphics[width=7.777cm]{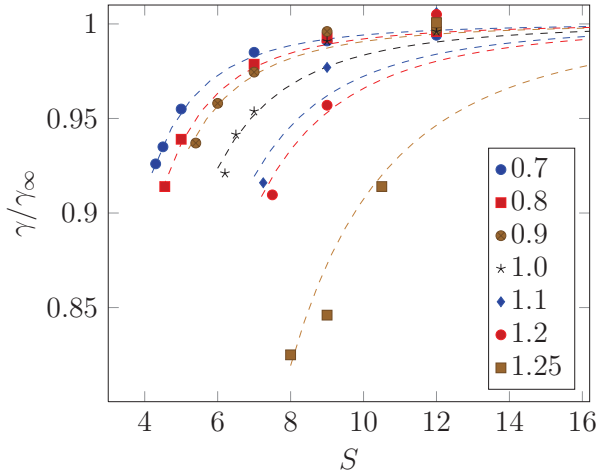}
\caption{Reduced surface tension $\gamma \slash \gamma_\infty$ over the slab thickness $S$ for different temperatures. The dashed lines represent the expression $\gamma \slash \gamma_\infty = 1 - a(T)/S^3$, where the temperature-dependent coefficients were adjusted to the simulation results, yielding $a(0.7) = 5.8$, $a(0.8) = 7.9$, $a(0.9) = 9.3$, $a(1.0) = 16$, $a(1.1) = 28$, $a(1.2) = 34$, and $a(1.25) = 93$.}
\label{fig:surfacetension_rrho}
\end{figure}

The differential pressure in the center of the liquid slab, i.e.\ the minimum differential pressure, is shown in Tabs.\ \ref{tab:data_pd1} and \ref{tab:data_pd2}. As already discussed, the differential pressure in the center of the liquid slab increases with decreasing slab thickness. The density corresponding to the minimum differential pressure, i.e.\ the maximum density, in the liquid slab is shown in Tabs.\ \ref{tab:data_rho1} and \ref{tab:data_rho2}. At the largest slab thickness the value agrees with the bulk properties. For lower slab thicknesses the density does not reach the bulk values. The minimum stable slab thickness increases with increasing temperature. Reducing the slab thickness below this point results in a rupture of the liquid phase and a transition from planar to cylindrical or spherical symmetry. 

\begin{figure}[t!]
\centering
 \includegraphics[width=7.777cm]{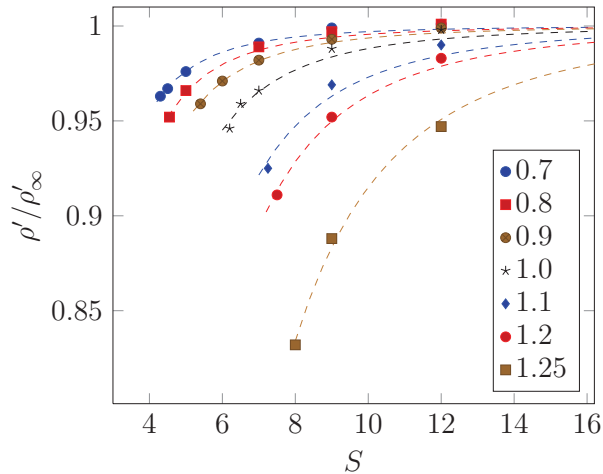}
\caption{Reduced density $\rho' \slash \rho'_\infty$ over the slab thickness $S$ for different temperatures. The dashed lines represent the expression $\rho' \slash \rho'_\infty = 1 - b(T)/S^3$, where the temperature-dependent coefficients were adjusted to the simulation results, yielding $b(0.7) = 3.0$, $b(0.8) = 4.4$, $b(0.9) = 6.3$, $b(1.0) = 12$, $b(1.1) = 27$, $b(1.2) = 37$, and $b(1.25) = 85$.}
\label{fig:density_rrho}
\end{figure}

In Fig.\ \ref{fig:surfacetension_rrho}, the relative surface tension -- reduced by $\gamma_\infty(T)$ as obtained from the large slab simulations -- is plotted over the slab thickness for different temperatures. Confinement between two planar vapor-liquid interfaces reduces the surface tension, and the numerical data suggest that this effect is of the order $1/S^3$.
In Fig.\ \ref{fig:density_rrho}, the reduced density $\rho' \slash \rho'_\infty$ is plotted over the slab thickness for different temperatures.
The relative density also decreases upon decreasing the slab thickness and, similar to the surface tension, this effect is approximately proportional to $1/S^3$ and becomes more significant at high temperatures.

For a slab thickness $S > 12$, the surface tension agrees with the value for a large liquid slab, within the simulation uncertainty. For a fluid described by the LJ potential, e.g.\ methane,\cite{VSH01}\ this means that confinement effects are significant for slabs which are thinner than 4.5 nm. At high temperatures, the density in the center of the liquid slab deviates more significantly from the bulk value.

\begin{figure}[h!]
\centering
 \includegraphics[width=7.777cm]{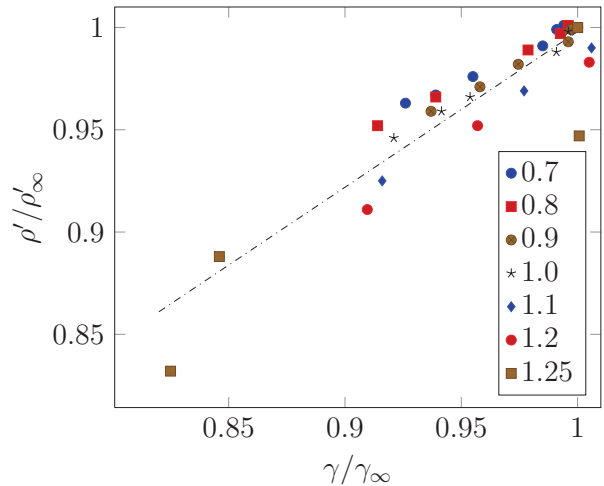}
\caption{Reduced density $\rho' / \rho'_\infty$ over the reduced surface tension $\gamma / \gamma_\infty$ for different temperatures. The dash dotted line represents the regression from Eq.\ (\ref{eq:regression}).}
\label{fig:density_surface}
\end{figure}

The present results suggest that the reduction of the density and the surface tension due to confinement are related effects. In Fig.\ \ref{fig:density_surface}, the respective ratios are displayed together, which discloses an approximately linear relationship. The regression 
\begin{equation}
 \frac{\rho'}{\rho'_\infty} \approx 0.76 \frac{\gamma}{\gamma_\infty} + 0.24
\label{eq:regression}
\end{equation}
is obtained from the simulation results.

\section{Conclusion}

\noindent
In the present work, molecular simulation was applied to study the influence of the slab thickness on the interfacial properties for planar vapor-liquid interfaces. The present results prove that such an effect exists for thin slabs and quantifies it for the LJ fluid. The surface tension decreases with decreasing slab thickness, and so does the density in the middle of the slab. The differential pressure does not reach zero for liquid slabs smaller than 12, which proves that under such conditions, a bulk-like region is absent. The confinement effects for the surface tension and the density were found to scale with $1/S^3$ in terms of the slab thickness $S$, so that a linear relation between both effects could be obtained.

The present results depart from those obtained by Weng et al.\ \cite{WPLT00} in a previous study, where no systematic correlation between the slab thickness and the surface tension was found. For a LJ system at $T$ = 0.818, Weng et al.\ detected minor fluctuations around a constant value ($\gamma$ = 0.78 $\pm$ 0.02), without a clear tendency, for a range of slab thicknesses between $S = 5.0$ and $9.0$.\cite{WPLT00}\ A juxtaposition with the present numerical data, cf.\ Tab.\ \ref{tab:data_gamma1} and Fig.\ \ref{fig:surfacetension_rrho}, according to which varying the slab thickness to such an extent has a significant influence on $\gamma$, clearly shows that there is a contradiction between present simulation results and the postulate of Weng et al.\ that ``with film thickness \dots{} surface tension values and density profiles show little variation.''\cite{WPLT00} For the simulations of Weng et al.,\cite{WPLT00}\ however, no long-range cutoff correction was employed at all, and the computations were only carried out over 120 000 time steps, as opposed to a million time steps for the present series of simulations. Since systems with an interface relax more slowly than the homogeneous bulk fluid, the extremely short simulation time could consitute a serious limitation, affecting the accuracy of the results obtained by Weng et al.\ \cite{WPLT00} to a significant extent. 

The comparison with results from a recent study of Malijevsk\'y and Jackson \cite{MJ12} suggests that the present results on confinement by two parallel planar vapor-liquid interfaces might also carry over qualitatively to confinement by the opposite sides of the single spherical interface that surrounds a small droplet. Therein, Malijevsk\'y and Jackson come to the conclusion that for liquid drops, the size dependence of the surface tension is best described by two distinct, additive terms: The conventional Tolman term, representing curvature, which increases the surface tension (i.e.\ the Tolman length is found to be negative), as well as ``an additional curvature dependence of the $1/R^3$ form'' which causes an eventual decrease of the surface tension ``for smaller drops.''\cite{MJ12}\ Furthermore, Malijevsk\'y and Jackson observe that the characteristic droplet radius, below which this negative corrective term becomes dominant, ``increases with increasing $r_\mathrm{c}$'' and conjecture that ``such a crossover occurs when \dots{} no `bulk' region can be assigned inside the drop. In this case even particles in the centre of the drop `feel' the interface.''

The present results lend further plausibility to this conjecture of Malijevsk\'y and Jackson. There could be a relation between their $1/R^3$ term and the $1/S^3$ confinement effect from the present study. According to such a hypo\-thesis, these contributions would both represent the deviation from bulk-like behavior of the liquid phase due to confinement.

\section*{Acknowledgment}

\noindent
The authors gratefully acknowledge financial support from Deutsche For\-schungsgemeinschaft (DFG) within the Collaborative Research Center (SFB) 926. They thank Stefan Eckelsbach, Philippe Ungerer, and Jadran Vrabec for fruitful discussions as well as Advait Deshpande and Srishti Srivastava from IIT Bombay for performing some of the present simulations. The present work was conducted under the auspices of the Boltzmann-Zuse Society of Computational Molecular Engineering (BZS), and the simulations of the largest slab were carried out on the XC 4000 supercomputer at the Steinbuch Centre for Computing, Karlsruhe, under the grant MOCOS.


\end{document}